\documentstyle[graphics,epsfig,floats,aps]{revtex}

\begin{document}

\draft
\catcode`\@=11 \catcode`\@=12
\twocolumn[\hsize\textwidth\columnwidth\hsize\csname@twocolumnfalse\endcsname
\title{ Co-ordination between Rashba spin-orbital interaction
and space charge effect and enhanced spin injection into
semiconductors}
\author{Wei Wu$^1$, Jinbin Li$^1$, Yue Yu$^1$ and S. T. Chui$^2$ }
\address{1. Institute of Theoretical Physics, Chinese Academy of
Sciences, P.O. Box 2735, Beijing 100080, China}
\address{2. Bartol Research Institute,
University of Delaware, Newark, DE 19716}

\date{\today}

\maketitle

\begin{abstract}
We consider the effect of the Rashba spin-orbital interaction and
space charge  in a
ferromagnet-insulator/semiconductor/insulator-ferromagnet junction
where the spin current is severely affected by the doping, band
structure and charge screening in the semiconductor. In diffusion
region, if the the resistance of the tunneling barriers is
comparable to the semiconductor resistance, the magnetoresistance
of this junction can be greatly enhanced  under appropriate doping
by the co-ordination between the Rashba effect and screened
Coulomb interaction  in the nonequilibrium transport processes
within Hartree approximation.
\end{abstract}

]

The spintronics is interesting since this involves exploration of
the extra degrees of freedom provided by electron spin, in
addition to those due to electron charge, which is believed to be
very useful in manipulating future electronic devices \cite{Piz}.
To realize such a spin device,  the Rashba spin-orbit interaction
is often considered \cite{Rashba}. Since this is caused by
structural inversion asymmetry in quantum wells, it can be
artificially controlled by adjusting the applied gate voltages and
specifically designing the heterostructure \cite{het}.

On the other hand, one of the current focus in spintronics lies in
injecting spin polarized electrons into non-magnetic
semiconductors\cite{Dat,Mon,Aro,STMST,mire,hu}. This is partially
motivated by the high magnetoresistance observed in ferromagnet
tunnel junctions\cite{moo}. However, in the diffusion region and
the room temperature, experiments have so far observed a small
magnetoresistance ratio of 1\% \cite{lee,ham} in
ferromagnet/semiconductor/ferromagnet structures. Rashba proposed
a ferromagnetic metal/tunneling-insulator/semiconductor (FIS)
junction to improve the spin injection rate \cite{Rash}. Many
different geometries of the tunneling junction were discussed in
Ref.\cite{fert}. It is possible that to lift the magnetoresistance
to near 10\% with a very high electric field \cite{fla}.

In a recent work \cite{Yue}, we investigated the space charge
effect in the nonequilibrium transport process within the Hatree
approximation and found that the magnetoresistance of a double FIS
junction could be greatly increased if one carefully adjusts the
parameters of the junction, such as the charge screening length
and the size of the semiconductor. The space charge effect in the
nonequilibrium transport process first was found playing an
important role in the characteristics of the device of
ferromagnetic/non-magnetic/ferromagnetic metal junction
\cite{chui}. Under steady state nonequilibrium conditions, a
magnetization dipole layer much larger than the charge dipole
layer is induced at the interface while the magnetization dipole
layer is zero under equilibrium conditions. We have applied this
idea to the double FIS junction \cite{Yue}.

We now ask the question: How do co-ordinations between the Rashba
spin-orbital interaction and the space charge affect the spin
transport in the double FIS junction? It was known that the
Coulomb interaction may enhance the Rashba effect. In a
one-dimensional Luttinger liquid formalism, Hausler has reported
an enhancement of the Rashba effect due to the spin charge
separation \cite{Haus}. Such an enhancement was also found in a
two-dimensional electron gas system \cite{chen}. However, the
effect of the Rashba spin-orbital interaction in the diffusion
region was thought to be not important. In this work, we would
like to investigate the co-ordination between the Rashba
spin-orbital interaction and the space charge effect in the double
FIS junction in the nonequilibrium transport process. To
completely solve the nonequilibrium problem with interactions is
highly non-trivial. It requires solving self consistently
Boltzmann type spin transport equation with the Poisson equation.
What we want to touch in this work is using a simple Hartree
approximation to check if the effect comes from these interactions
is significant. Consequently, it is found that if (i) the
resistance of the tunneling barrier is comparable to the
semiconductor resistance and (ii) the $n$-type semiconductor has
an appropriate doping, then while the magnetoresistance of the
junction is greatly enhanced as the charge screening length
becomes shorter as we already showed in \cite{Yue}, it is grown up
as the Rashba spin-orbital coupling. The shorter charge screening
length is , the more gain in the magnetoresistance comes from the
Rashba term. Comparing with the non-interacting model where the
Rashba effect may not be observed in the diffusion region, we see
a great increment of the magnetoresistance comes from the
co-ordination between the Rashba spin-orbital interaction and
space charge effect.

\vspace{2mm}

\noindent{\it Rashba Spin-Orbital Coupling:} We consider a
two-dimensional electron gas in a narrow gap quantum well, such as
those based on InAs. The spin-orbital coupling in this kind of
systems is dominated by Rashba term \cite{lomm}:
\begin{eqnarray}
H_{so}=\alpha(\sigma_xp_y-\sigma_yp_x), \label{so}
\end{eqnarray}
where $\sigma_{x,y}$ are the Pauli matrices and the Rashba
parameter $\alpha$  is determined by the asymmetry of the
potential confining the electrons in the two-dimensional $x$-$y$
plan and can be controlled by a gate voltage \cite{het}. For the
system we are considering, its value is about $10^{-12}-10^{-11}$
eVm \cite{alpha}. For a double FIS junction where the current
flows in the $x$-direction, we can estimate the spin dependence of
the density of state and the resistance of the electron gas.
According to the Rashba term (\ref{so}), the single particle
dispersion reads
\begin{eqnarray}
E_\sigma=\frac{\hbar^2k^2_\sigma}{2m^*}+\sigma\alpha k_\sigma,
\end{eqnarray}
where $m^*\approx 0.04m_e$ for the bare electron mass $m_e$ and
$k_\sigma=k+\sigma\frac{m^*\alpha}{\hbar^2}$; $\sigma$ is called
spin-orbital coupling label \cite{Feve}. The Rashba spin-ortial
interaction contributes a current $j_{so}$ to the system. However,
in the diffusion region, such an effect may be small \cite{Feve}.
Here, we would like to see another effect arising from the Rashba
term. When the Rashba term is non-zero, it is easy to see the
density of state $N^{so}_{N}$ is given by \cite{Feve}
\begin{eqnarray}
N^{so}_{N}= N^N_{F}(1+\alpha_N),
\end{eqnarray}
where $N^N_{F}$ is the density of state at the Fermi surface of
the semiconductor for $\alpha=0$ and $\alpha_N=\frac{\alpha}{\hbar
v_F}$ for the Fermi velocity $v_F$. It is seen that the density of
state is not $\sigma$-dependent \cite{Feve}. For a typical
two-dimensional electron gas, the electron density is of order
$\rho_e=10^{11}$ and $10^{12}$ cm$^{-2}$. Thus, using
$v_F=\frac{k_F}{m^*}=\frac{\hbar}{m^*} \sqrt{2\pi \rho_e}$,
$\alpha_N$ is in order of $10^{-2}$ and $10^{-3}$.  In the
following, we will see that the magnetoresisatnce of the double
FIS junction is very sensitive to the change of the charge
screening length which is determined by the density of state.

\vspace{2mm}

{\noindent \it Description of the Junction: } In the diffusion
region and the room temperature, we consider the junction
F1-I1-S-I2-F2 along the $x$-direction. The thickness of the metal
(F1, F2), the insulating barriers (I1, I2) and the semiconductor
(S) are denoted by $L^{L,R}$, $d_{1,2}$ and $x_0$, respectively
(See, Figure 1). For the practical case, $x_0$ is less than the
spin diffusion length $l_N$ of the two-dimensional electron gas in
the semiconductor. The charge screening lengths in the metal and
the semiconductor are denoted by $\lambda_{L,R}$ and $\lambda_N$,
respectively. In our model, we assume $\lambda_N\ll l_N$ and
$\lambda_{R,L}\ll l_{R,L}$, the spin diffusion lengths in the
metal. Typically, $\lambda\sim 10^{-1}$nm, $l\sim 10^1$nm, and
$l_N> 1\mu$m. $x_0\sim 100$nm to $1\mu$m, depending on the
structure of junctions. The screening length in the semiconductor,
$\lambda_N$, is dependent on the doping of the semiconductor and
can vary in a wide range, say 10nm for the heavy doped
semiconductor and 100nm-1$\mu$m for the lightly doped or undoped
semiconductor. To avoid discussing the spin-orbit splitting of the
heavy and the light-holed bands near the zone center, we consider
the $n$-type semiconductor only.

The problem we would like to solve has been defined in
\cite{chui,Yue}. Summarily, it is four sets of equations

\vspace{1mm}

\noindent (1)  The total charge-current conservation is described
by
\begin{eqnarray}
\nabla\cdot {\bf j}=-\frac{\partial \rho}{\partial t},\label{1}
\end{eqnarray}
where $\rho$ is the charge density.

\vspace{1mm}

\noindent (2) The spin $s$-dependent current is determined by the
diffusion equations. In order to allow an analytic analysis, we
take a simple Hartree approximation into account to see the space
charge effect. Under such an approximation, the diffusion
equations reads \cite{chui}
\begin{eqnarray}
j_s=\sigma_s(\nabla\mu_s-\nabla W_0+E),\label{2}
\end{eqnarray}
where the magnitude of the electric charge has been set as one;
$E$ is the external electric field; the chemical potential $\mu_s$
is related to the charge density $\rho_s$ by $\nabla\mu_s
=\frac{\nabla\rho_s}{N_s}$ where $N_s$ is the spin-dependent
density of state. $W_0=\int d\vec r' U_{int}(\vec r-\vec r'
)\rho(\vec r') $ is the potential caused by the screened Coulomb
interaction $U_{int}(\vec r)$ and $\nabla W_0$ is the so-called
screening field induced by $U_{int}(\vec r)$.

\vspace{1mm}

\noindent (3) The third is the magnetization relaxation equation
where the magnetization density $M=\rho_\uparrow-\rho_\downarrow$
relaxes with a renormalized spin diffusion length $l$. In the
relaxation time approximation, one has
\begin{eqnarray}
\nabla^2M-M/l^2=0.\label{3}
\end{eqnarray}
(4) The boundary conditions at the interfaces are given by
\begin{eqnarray}
\Delta\tilde\mu_s-\Delta W=r(1-s\gamma)j_s, \label{4}
\end{eqnarray}
where $\tilde\mu_s=\mu_s+Ex$; $Ex$ is the voltage drop on the left
side of the barrier and $\Delta W $ is the electric potential drop
across the barrier, which is assumed much smaller than
$\Delta\mu_s$; $r_s=r(1-s\gamma)$ is the barrier resistance. We
assume that there is no spin relaxation in the insulator, the
spin-dependent currents are continuous across the junctions,
$j^L_s(-d_1/2)=j^N_s(d_1/2)$ and
$j^R_s(x_0+d_2/2)=j^N_s(x_0-d_2/2)$.

In addition, we have the neutrality condition for the total
charges ($Q_{L,N,R}$, e.g., $Q_N=\int^{x_0+d_1/2}_{d_1/2} \rho
dx$) accumulated at the interfaces. By Gauss' law, for the point
$d_1/2<x<x_0+d_1/2$, i.e., inside the semiconductor, the potential
$W_0$ is determined by
\begin{eqnarray}
\nabla W_0(x)= 4\pi Q_L+4\pi\int_{d_1/2}^x\rho dx, \label{5}
\end{eqnarray}
whose constant part of the right hand side gives the constraint on
the charge while the $x$-dependent part gives the function form of
the potential $W_0$. Another constraint is the neutrality of the
system:
\begin{eqnarray}
Q_L+Q_N+Q_R=0. \label{6}
\end{eqnarray}

With those sets of equations (Eqs.(\ref{1})-(\ref{4})) and the two
constraint on the charges (Eqs.(\ref{5}) and (\ref{6})), the
problem can be solved. The formal solutions of the problem are
\begin{eqnarray}
\rho^{L}(x)&=&\frac{\lambda_L}{l_L}\rho^L_{10}e^{(x+d_1/2)/\lambda_L} +\frac{%
\lambda_L^2}{l_L^2}\rho^L_{20}e^{(x+d_1/2)/l_L},  \nonumber \\
M^L(x)&=&M^L_0(1-\frac{\lambda_L^2}{l_L^2})e^{(x+d_1/2)/l_L},\label{metal}
\end{eqnarray}
with similar solutions for the right hand side. In the
semiconductor, if $\lambda_N\ll x_0$,
\begin{eqnarray}
\rho^N(x)&=&\rho^{(1)}(x)+\rho^{(2)}(x),  \nonumber \\
\rho^{(1)}(x)&=&\frac{\lambda_N}{l_N}\rho^{(1)}_{10}e^{-(x-d_1/2)/\lambda_N} +\frac{%
\lambda_N^2}{l_N^2}\rho^{(1)}_{20}e^{-(x-d_1/2)/l_N},  \nonumber
\\
\rho^{(2)}(x)&=&\frac{\lambda_N}{l_N}\rho^{(2)}_{10}e^{(x-x_0+d_2/2)/\lambda_N}
\nonumber \\
&+&\frac{\lambda_N^2}{l_N^2}\rho^{(2)}_{20}e^{(x-x_0+d_2/2)/l_N}.\label{semi}
\end{eqnarray}
$M^{(1),(2)}$ can  be obtained similarly. All of coefficients in
(\ref{metal}) and (\ref{semi}) can be determined by using
Eqs.(\ref{1})-(\ref{4}) and the constraint (\ref{5}) and
(\ref{6}). The screening potential $W_0$ is determined by Gauss'
law. The total current is $j=\sum_s j_s$. Although $j_s$ are not a
constant, the total current $j$ is still a constant.

 \vspace{2mm}

\noindent{\it The spin-dependent currents}: It is necessary to
simplify the problem to demonstrate the essential physics. One
sets the parameters of the metals and barrier widths on the left
and right sides to be the same: $\lambda_R=\lambda_L=\lambda$,
$l_L=l_R=l$, $d_1=d_2=d$ and so on. The resistances of the barrier
layers are taken as $r^{(1)}=r^{(2)}=r$;
$\gamma_1=\gamma_2=\gamma$ for the parallel configuration and
$\gamma_1=-\gamma_2=\gamma$ for the anti-parallel configuration.
To illustrate, we focus on the calculation in the left barrier
located at $x=0$. The tunneling resistance is given by
\begin{equation} r_s^{(1)}=r(1-\gamma s)=
r_{0,s}\exp[d(\kappa_s(\mu)-\kappa_s(0))],
\end{equation}
where $
\kappa_s(\mu)\propto\int_0^ddx[2m(U-\Delta\mu_s(0)x/d)]^{1/2}, $
with $U$ the barrier height. The current  $j^L_s(x)$ at $x=0$ is
dependent on the bias voltage and the interaction, which is given
by
\begin{eqnarray}
j_s^L(0)= A_sj_{0s},\label{current}
\end{eqnarray}
where, for the ferromagnet metal on left hand side,
\begin{eqnarray}
A_s=1+\frac{4\pi\lambda^2}{l}\frac{\alpha(\beta-s)}{1-\alpha\beta}
\frac{\rho_{10}^L+M_0^L}{E},
\end{eqnarray}
and $j_{0s}=\sigma_s E$ is the current with no interaction and the
current $j_{so}$ that is contributed to from the Rashba term has
been neglected; $\beta$ ($\alpha$) measures the spin asymmetry of
the conductivities $\sigma_s$ (densities of states at the Fermi
surface $N_s$): $\sigma_s=\frac{\sigma}{2}(1+\beta s)$ and
$N_s=\frac 12N_F(1+s\alpha )$ where $N_F$ is related to the
screening length $\lambda$ by $\frac 1{N_F}=2\pi \lambda
^2\frac{1-\alpha ^2}{1-\alpha \beta }$. Noting that both
$\rho^L_{10}$ and  $M_0^L$ are proportional to the external
electric field $E$, $A_s$ is solely determined by the material
parameters.

Eq. (\ref{current}) implies that the spin-dependent current
$j_s(0)$ passing the interface differs a factor $A_s$ from the
non-interacting current $j_{0s}(0)$. For the parallel
configuration, $j_{0s}(0)$ is given by \cite{Yue}
\begin{eqnarray}
j^{s}_{0s}(0)\approx \frac{ V}{R^{so}_{N}+2r_{0,s}Y_s(0)},
\end{eqnarray}
where $ Y_s(0)=e^{\kappa_{0s
}d[\frac{2}{3\Delta\hat\mu_s(0)}(1-(1-\Delta\hat\mu_s(0))^{3/2})-1]}$
and  $R^{so}_{N}= R_N(1-\beta_N)$ with $\beta_N\approx\alpha_N$.
For the anti-parallel configuration,
\begin{eqnarray}
&&j^a_{0s}(0)\approx\frac{V}{R^{so}_{N}+\sum_sr_{0,s}Y_s(0)}.
\end{eqnarray}

\vspace{2mm}

{\noindent \it The magnetoresistance:} Since the total current is
constant everywhere, we have, for the parallel configuration $
\frac{1}{R_P}=\sum_sj_s(0)/V $ and for the anti-parallel
configuration, $ \frac{1}{R_{AP}}=\sum_sj_s(0)/V $. From these, we
obtain the magnetoresistance ratio
\begin{eqnarray}
\frac{\Delta
R}{R}\equiv\frac{R_{AP}-R_P}{(R_{AP}+R_P)}=\frac{X}{2+X},\label{int}
\end{eqnarray}
where $X=\sum_s\frac{A^P_s(R^{so}_{N}+\sum_{s'}r_{0,s'}
Y^{AP}_{s'}(0))} {2(R^{so}_{N}+2r_{0,s}Y^P_s(0))}-1$.  For
non-interacting electrons, $A^P_+=A^P_-=1$ and we knew that that
$\Delta R/R$ will not be beyond a maximal value (at $V\to 0$)
about 3.2\% for $r_{0+}:r_{0-}:R_N=1:2:1$ and decays as the bias
voltage increases \cite{Yue}. Since $j_{so}$ has been neglected,
there is no observable Rashba effect without interactions.  After
the interaction is included, the ratio (\ref{int}) grows greatly
and the Rashba effect is enhanced as $\lambda_N$ becomes smaller.
This can be clearly seen in Figs. 2 and 3. We take $x_0=1.25\mu$m,
$\lambda=0.1$nm, $l=20$nm, and $l_N=3\mu$m and set
$\alpha=\beta=1/2$. The different choice of $\alpha$ and $\beta$
will not qualitatively affect the result if they do not deviate
from 1/2 too much.  In Fig. 2, we depict the magnetoresistance
versus the bias voltage for $\lambda_N=100$nm with $\alpha_N$ from
0 to 0.05. It is seen that $\Delta R/R$ raises from 16\% to 18\%
when $\alpha_N$ from 0 to 0.05. In Fig. 3, for $\lambda_N=50$nm,
it is shown that the Rashba effect lifts much faster. $ \Delta
R/R$ raises from 25\% to 30\% for $\alpha_N$ from 0 to 0.05.
Hence, we see a strong co-ordination between the Rashba
spin-orbital and the screened Coulomb interactions to increase the
magnetoresistance. While the Coulomb interaction largely enhances
the magnetoresistance, the Rashba spin-orbital interaction may
enhance $\Delta R/R$. This kind of Rashba effect becomes more
significant in a shorter charge screening length.

In conclusions, we have shown the co-ordination between the Rashba
spin-orbital and screened Coulomb interactions on electron
injection from ferromagnet to semiconductor. The magnetoresistance
grows fast as the charge screened length in the semiconductor
becomes shorter.  The Rashba term also enhances the
magnetoresistance and plays more important role as the Coulomb
interaction is stronger. In fact, by a chose examination to our
solution, this Rashba effect is corresponding to a renormalization
to the screening length $\lambda_N\to
\lambda^*_N=\lambda_N/(1+\alpha_N)$ in the dominate terms of the
solution. Because our solution is very sensitive to $\lambda_N$,
it is understood that why a small spin-orbital coupling can cause
a relative large gain in the magnetoresistance.

This work was partially supported by the NSF of China.

\vspace{2mm}

\noindent Fig. 1 The sketch of the double FIS junction.

\vspace{2mm}

\noindent Fig. 2 The magnetoresistance versus the bias voltage (in
unit $U$) with $\lambda_N=100$nm. The junction parameters are
$x_0=1.25\mu$m, $\lambda=0.1$nm, $l=20$nm,  and $l_N=3\mu$m. Three
different values of $\alpha_N$ are taken.

\vspace{2mm}

\noindent Fig. 3 The magnetoresistance versus the bias voltage
with $\lambda_N=50$nm. Other parameters are the same as those in
Fig. 2.


\begin{references}

\bibitem{Piz} G. A. Prinz, Phys. Today {\bf 48}, No. 4, 58 (1995);
G. A. Prinz, Science {\bf 282}, 1660 (1998). S. A. Wolf, D. D.
Awschalom, R. A. Buhrman, J. M. Daughton,5 S. von Molnar, M. L.
Roukes, A. Y. Chtchelkanova, D. M. Treger, Science {\bf 294}, 1492
(2001).

\bibitem{Rashba} E. I. Rashba, Sov. Phys. Solid State, {\bf 2},
1109 (1960).

\bibitem{het} J. Nitta, T. Akazaki, H. Takayanagi, and T. Enoki,
Phys. Rev. Lett. {\bf 78}, 1335 (1997).  G. Engels, J. Lange, Th.
Schapers, and H. Luth, Phys. Rev. B {\bf 55}, R1958 (1997). Th.
Schapters, G. Engels, J. Lange, Th. Klocke, M. Hollfelder, and H.
Luth, J. Appl. Phys. {\bf 83}, 4324 (1998). D. Grundler, Phys.
Rev. Lett. {\bf 84}, 6074 (2000). G. L. Chen, J. Han, T. T. Huang,
S. Datta, and D. B. Janes, Phys. Rev. B {\bf 47}, 4084 (1993).




\bibitem{Dat} S. Datta and B. Das, Appl. Phys. Lett. {\bf 56}, 665
(1990).

\bibitem{Mon} F. G. Monzon, M. Johnson and M. L. Roukes, Appl.
Phys. Lett. {\bf 71}, 3087 (1997).

\bibitem{Aro} A. G. Aronov and G. E. Pikus, Sov. Phys. Semicond.
{\bf 10}, 698 (1976).




\bibitem{STMST} S. F. Alvarado and P. Renaud, Phys. Rev. Lett. {\bf 68},
1387 (1992); V. P. LaBella, D. W. Bullock, Z. Ding, C. Emergy, A.
Venkatesen, W. F. Oliver, G. J. Salamo, P. M. Tibado and M.
Mortazavi, Science {\bf 292}, 1518 (2001); H. J. Zhu, M.
Ramsteiner, H. Kostial, M. Wassermeier, H. P. Sch\"onherr and K.
H. Ploog, Phys. Rev. Lett. {\bf 87}, 16601 (2001); A. T. Hanbicki,
B. T. Jonker, G. Itskos, G. Kioseoglou and A. Petrou, Appl. Phys.
Lett. {\bf 80}, 1240(2002).

\bibitem{mire} F. Mireles and G. Kirczenow, e-Print:
cond-mat/0206009, accepted for publication by Phys. Rev. B.


\bibitem{hu} C.-M. Hu and T. Matsuyama, Phys. Rev. Lett. {\bf 87},
066803 (2001).

\bibitem{moo} J. S. Moodera, Lisa R. Kinder, Terrilyn M. Wong and
R. Meservey, Phys. Rev. Lett. {\bf 74}, 3273 (1995); J. S. Moonera
and Lisa R. Kinder, J. Appl. Phys. {\bf 79}, 4724 (1996).

\bibitem{lee} W. Y. Lee, S. Gardelis, B. C. Choi, Y. B. Xu, C. G.
Smith, C. H. W. Barnes, D. A. Ritchie, E. H. Linfield, and J. A.
Bland, J. Appl. Phys. {\bf 85}, 6682 (1999).

\bibitem{ham} P. R. Hammar, B. R. Bennet, M. J. Yang, and M.
Johnson, Phys. Rev. Lett. {\bf 83}, 203 (1999).

\bibitem{Schm} G. Schmidt, D. Ferrand and L. W. Molenkamp, Phys.
Rev. B, {\bf 62}, R4790 (2000).

\bibitem{grun} D. Grundler, Phys. Rev. B {\bf 63}, 161307 (2001).


\bibitem{Rash} E. I. Rashba, Phys. Rev. B {\bf 62}, R16267 (2000).
The update progress in the F-I-S-I-F junction, see, E. I. Rashba
cond-mat/0206129.

\bibitem{fert} A. Fert and H. Jaffres, Phys. Rev. B {\bf 64},
184420 (2001).

\bibitem{fla} Z. G. Yu and Flatte, Phys. Rev. B {\bf 66}, 201202
(2002).

\bibitem{Yue} Yue Yu, Jinbin Li and S. T. Chui, cond-mat/0212133.

\bibitem{chui} S. T. Chui, Phys. Rev. B {\bf 52}, R3882 (1995); J.
Appl. Phys. {\bf 80}, 1002 (1996); Phys. Rev. {\bf 55}, 5600
(1997). S. T. Chui and J. Cullen, Phys. Rev. Lett. {\bf 74}, 2118
(1995). S. T. Chui, Phys. Rev. B {\bf 55}, 5600 (1997); S. T.
Chui, J.-T. Wang, L. Zhou, K. Esfarjani and Y. Kawazoe, J. Phys.
Conds. Matt. {\bf 13}, L49 (2001); S. T. Chui and L-. B. Hu, Appl.
Phys. Lett. {\bf 80}, 273 (2002).

\bibitem{Haus} W. Hausler, Phys. Rev. B {\bf 63}, 121310 (2001).

\bibitem{chen} Guan-Hong Chen and M. E. Raikh, Phys. Rev. B {\bf
60}, 4826 (1999).

\bibitem{lomm} G. Lommer, F. Malcher, and U. Rossler, Phys. Rev.
Lett. {\bf 60}, 728 (1988).

\bibitem{alpha} See, for examples, B. Das, S. Datta, and
Reifenberger, Phys. Rev. B {\bf 41}, 8278 (1990);J. Luo, H.
Munekata, F. F. Fang, and P. J. Stiles, Phys. Rev. B {\bf 41},
7685 (1990); J. P. Heida, B. J. van Wess, J. J. Kuipers, T. M.
Klapwijk, and G. Borghs, Phys. Rev. B {\bf 57}, 11911 (1998); J.
Nitta et al in \cite{het}; T. Koga, J. Nitta, T. Ahazaki, and H.
Takayanagi, Phys. Rev. Lett. {\bf 89}, 046801 (2002).

\bibitem{Feve} G. Feve, W. D. Oliver, M. Aranzana, and Y. Yamamoto
Phys. Rev. B {\bf 66}, 155328 (2002).


\end{references}
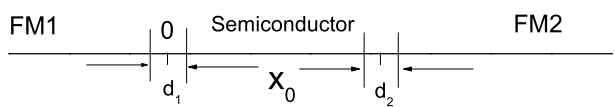
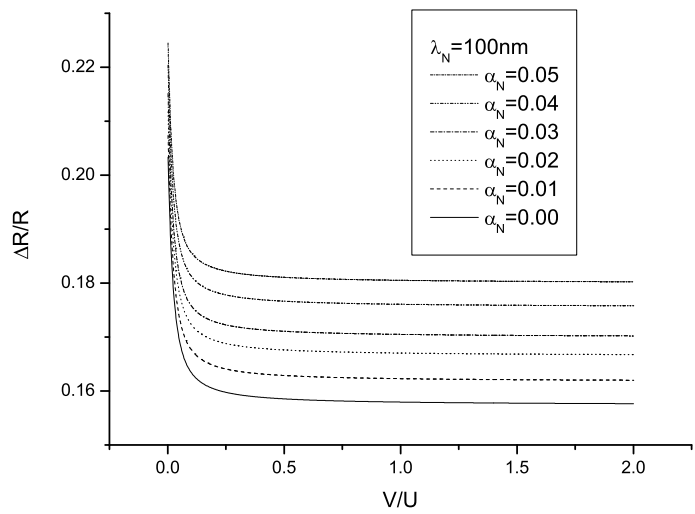
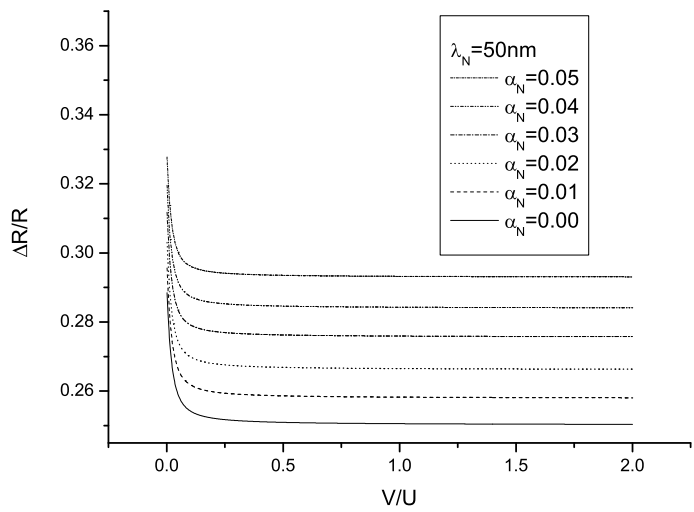
\end{document}